# Spectrum Cascade Bloch Oscillations in Temporally Modulated Acoustics

Chengzhi Qin,[1,†], Yugui Peng,[1,2,†], Ying Li[2], Xuefeng Zhu,[1,*], Bing Wang,[1,*], Cheng-Wei Qiu[2,*] and Peixiang Lu[1,3,*]

[1]School of Physics and Wuhan National Laboratory for Optoelectronics, Huazhong University of Science and Technology, Wuhan, 430074, China.

[2]Department of Electrical and Computer Engineering, National University of Singapore, Singapore 119620, Singapore.

[3]Laboratory for Optical Information Technology, Wuhan Institute of Technology, Wuhan, 430205, China.

E-mails: xfzhu@hust.edu.cn; wangbing@hust.edu.cn; eleqc@nus.edu.sg; lupeixiang@hust.edu.cn.

Abstract

Bloch oscillations (BOs) refer to a periodically oscillatory motion of particle in lattice systems driven by a constant force. By temporally modulating acoustic waveguides, BOs can be generalized from spatial to frequency domain, opening new possibilities for spectrum manipulations. The modulation can induce mode transitions in the waveguide band and form an artificial frequency lattice, with the mismatched wave vector during transitions acting as a constant force that drives frequency Bloch oscillations (FBOs). Furthermore, the modulation phase accompanying transitions serves as a gauge potential that controls the initial oscillation phase, providing an additional degree of freedom to tailor FBOs. We report that multiple FBOs with judiciously designed oscillation phases can be further cascaded to realize acoustic spectrum self-imaging, unidirectional transduction and bandwidth engineering. The study proposes the concept of FBOs in acoustic systems and functionalizes its cascade configurations for advanced control of sound spectrum. This paradigm may find versatile applications in underwater secure communication, voice encryption and signal processing.

Acoustic wave has found its great importance in a variety of wave physics researches and practical applications [1-3]. Controlling the spatial degrees of freedom of acoustic waves, such as using sonic crystals, metamaterials and metasurfaces, has received intensive attention. Examples include acoustic imaging [4-11], vortex [12-16] and accelerating beams [17-20], unidirectional diffraction [21-24] and topological insulators [25-30]. Nevertheless, these studies mainly concern static acoustic systems, in which the frequency or energy is a conserved quantity. The research of manipulating sound frequency is still in its infancy, while it is highly desirable in versatile applications such as acoustic communication, information encryption and processing [13]. Previously, the frequency conversion was achieved via acoustic nonlinear effect, namely, the second harmonic wave generation [31-33]. However, the conversion efficiency is quite low even for high-intensity acoustic waves. Alternatively, spectrum shift for acoustic waves has also been realized through dynamic scattering from rotating objects [34]. Since this approach depends sensitively on the shape and orientation of the scatterers, it is still far from implementing the precise control of sound spectrum.

According to the energy conservation law, the frequency of a system can be changed by applying time-variant perturbations. In this Letter, we propose an acoustic waveguide system with temporal modulation of bulk modulus, so as to manipulate the spectrum and energy of sound. In the presence of the modulation-induced frequency lattice, the wave vector mismatch during transitions acts as an effective constant force that induces frequency Bloch oscillations (FBOs). The initial phase of time modulation mimics an effective gauge potential that controls the initial oscillation phase of FBOs. By driving two cascade FBOs with different oscillation phases in two separated modulated waveguides, we demonstrate a prototype of acoustic secure communication based on the spectrum self-imaging effect. By cascading multiple waveguides with out-of-phase time modulations, we also break the intrinsic localized feature of FBOs and realize spectral unidirectional transduction and bandwidth manipulation. Our study reveals the capability of cascade FBOs in the manipulation of acoustic spectrum for a plethora of important applications.

We start from revealing the effect of FBOs induced in one time-modulation acoustic waveguide. The waveguide is filled with silicone rubber [35-37], a water-like ultrasonic material. Suppose the compressibility is subject to a time modulation $\beta(t)=\beta_0+\Delta\beta\cos(\Omega t+\phi)$, where $\beta_0$, $\Delta\beta$, $\Omega$ and $\phi$ are the background compressibility, modulation amplitude, frequency, and initial phase, respectively. In practice, as shown in Fig. 1(a), the modulation can be realized by compressing the filling medium through piezoelectric actuators [36, 37]. The general acoustic wave equations under time modulation are given by [3, 36]

$$\begin{cases} \dfrac{\partial \rho(\mathbf{r},t)}{\partial t} = -\rho_0 \nabla \cdot \mathbf{v}(\mathbf{r},t), \\ \dfrac{\partial \mathbf{v}(\mathbf{r},t)}{\partial t} = -\dfrac{\nabla p(\mathbf{r},t)}{\rho_0}, \\ \dfrac{\partial \rho(\mathbf{r},t)}{\partial t} = \rho_0 \dfrac{\partial [\beta(t) p(\mathbf{r},t)]}{\partial t}, \end{cases} \quad (1)$$

where $\rho(\mathbf{r}, t)$, $p(\mathbf{r}, t)$ and $\mathbf{v}(\mathbf{r}, t)$ are the medium mass density, pressure, and particle velocity. $\rho_0$ is the time-independent background mass density. The governing equation for the acoustic wave thus reads

$$\nabla^2 p(\mathbf{r},t) - \rho_0 \frac{\partial^2}{\partial t^2}[\beta(t) p(\mathbf{r},t)] = 0, \quad (2)$$

For simplicity, we consider the one-dimensional case with $\nabla^2 = \partial^2/\partial z^2$ and $\mathbf{r}$ replaced by $z$. As shown in Fig. 1(b), the time-periodic modulation can induce multiple intraband transitions in the linear waveguide band and create an artificial frequency lattice with interval $\Omega$. The instantaneous pressure field is thus $p(z, t) = \sum_n p_n(z) \exp[i(\omega_n t - k_n z)]$, where $\omega_n = \omega_0 + n\Omega$ and $k_n = k_0 + n\Delta k$ ($n = 0, \pm 1, \pm 2, \cdots$) are the angular frequency and wave number of $n$th-order fundamental mode. Substituting $p(z,t)$ into Eq. (2), we can derive the coupled-mode equation [38]

$$i\frac{\partial p_n(z)}{\partial z} = C_n \left( e^{i(\phi + \Delta k z)} p_{n-1}(z) + e^{-i(\phi + \Delta k z)} p_{n+1}(z) \right), \quad (3)$$

where $C_n = C_0 = k_0(\Delta\beta/4\beta_0)$ is the coupling strength between adjacent modes in the frequency lattice.

To understand the spectrum evolution, we consider the eigen Bloch mode in the frequency lattice, namely, an infinite-width frequency comb with $p_n(z) = p_0 \exp(in\phi_0) \exp(ik_z z)$, where $p_0$ is the uniform amplitude, $\phi_0$ the Bloch momentum in the frequency dimension and $k_z$ the collective propagation constant along the waveguide direction. Plugging the Bloch mode into Eq. (3), we can obtain the lattice effective band structure

$$k_z[\phi_\omega(z)] = -2C \cos[\phi_\omega(z) - \phi], \quad (4)$$

where $\phi_\omega(z) = \phi_0 - \Delta k z$ is the $z$-dependent Bloch momentum. Since $z = c_0 t$, the Bloch momentum also varies in time $\phi_\omega(t) = \phi_0 - \Delta k c_0 t$, where $c_0$ is the speed of sound in the medium. As denoted in Fig. 1(c), the time-varying Bloch momentum corresponds to an effective force $F_{eff} = \partial \phi_\omega(t)/\partial t = -\Delta k c_0$, which stems from the wave vector mismatch during frequency transitions. Here $\phi_\omega(z)$ is shifted by $\phi$ in the band structure, indicating that the modulation phase $\phi$ plays the role of an effective gauge potential through $\int_{\omega_n}^{\omega_{n+1}} A_{eff} d\omega = \phi$ or equivalently $A_{eff}=\phi/\Omega$. Since $\phi_\omega(z)=k_\omega(z)\Omega$, the band structure can be rewritten as $k_z[k_\omega(z)]=-2C\cos[(k_\omega(z)-A_{eff})\Omega]$. The band structure shift induced by the gauge potential is analogous to the situation where the kinetic momentum of an electron is substituted by the conical momentum in the presence of an electromagnetic vector potential [39, 40]. For the input of a real finite-width Bloch wave packet centered at an initial Bloch momentum $\phi_0$, the group velocity in the frequency lattice is $z$-periodic, $v_g(z)=-\partial k_z[k_\omega(z)]/\partial k_\omega(z)=-2C\Omega\cos(\phi_0-\Delta kz-\phi)$, which gives rise to a frequency shift $\omega(z) = \omega(0) + \int_0^z v_{g,\omega}(z')dz'$

$$\omega(z) = \omega(0) + \frac{2C\Omega}{\Delta k}[\cos(\phi-\phi_0) - \cos(\Delta kz + \phi - \phi_0)], \tag{5}$$

where $\omega(0)$ denotes the initial central frequency. The center-of-mass of the wave packet undergoes an oscillatory motion in the frequency lattice, which is termed frequency Bloch oscillations (FBOs). The corresponding oscillation period is $Z_B=2\pi/|\Delta k|$, which is inversely proportional to the magnitude of the driving force. Since $\omega(z)=\omega(z+vZ_B)$, ($v=1, 2, \cdots$), the frequency comb manifests self-imaging at integer multiples of oscillation period $Z_B$. Furthermore, we point out that the initial oscillation phase is determined by the gauge potential, namely, the initial phase of time modulation. In the following, we will exploit this gauge degree of freedom by cascading multiple FBOs with different oscillation phases, aiming to extend the functionalities of acoustic spectrum manipulation.

Firstly, we consider a configuration comprising two modulation waveguides separated by an unmodulated waveguide channel. We find this system has the potential for acoustic long-range, secure communication where the information encoded in the frequency dimension can be perfectly reconstructed at the output port. As shown in Fig. 2(a), the two waveguides modulated with the same frequency $\Omega$ but different phases ($\phi_1$ and $\phi_2$) can function as the encoding and decoding parts in an acoustic communication system, with the central unmodulated waveguide serving as the transmission channel. The question lies in how to design the two modulation waveguide lengths and modulation phases such that an arbitrarily input spectrum can be perfectly reconstructed after passing through the whole system. As shown in Fig. 2(b), we consider a Bloch-mode wave packet with initial Bloch momentum $\phi_0$ is incident into the system, it manifests FBOs (represented by the dashed red curve) in the encoding waveguide such that $\phi_\omega(z)=\phi_0-\Delta kz$, which reaches $\phi_0-\Delta kL_1$ in its end. In the unmodulated waveguide channel, though no FBOs occur, the phase difference between adjacent order modes still accumulates due to the presence of waveguide dispersion (denoted by the green arrows). Here we assume the channel waveguide length is $L_p$, the phase difference thus evolves into $\phi_0 - \Delta k L_1 + \Delta k L_p$ in its end. When entering the decoding waveguide, FBOs restart and the Bloch momentum again manifests a linear shift with $\phi_\omega(z) = (\phi_0 - \Delta k z + \Delta k L_p) - \Delta k(z - L_1 - L_p) = \phi_0 + 2\Delta k L_p - \Delta k z$ (represented by the blue dashed curve). Due to the distinct gauge potentials in the two modulation waveguides, the Bloch momentum undergoes different shifts, as if in two different phase reference frames of $\phi_1$ and $\phi_2$. So we can obtain the relative Bloch momentum $\phi_\omega(z) - \phi_m$ ($m = 1, 2$) in the two waveguides

$$\phi_{\omega,m}(z) = \begin{cases} (\phi_0 - \Delta k z) - \phi_1, & (m = 1) \\ (\phi_0 + 2\Delta k L_p - \Delta k z) - \phi_2, & (m = 2) \end{cases} \quad (6)$$

To realize the spectrum reconstruction at the output port of the whole system, two conditions must be simultaneously satisfied. Condition 1: The Bloch momentum at the output port of the encoding waveguide should equal that at the input port of the decoding waveguide, namely, $\phi_{\omega,1}(L_1) = \phi_{\omega,2}(L_1 + L_p)$. Condition 2: The total lengths of encoding and decoding waveguides should be integer multiples of the Bloch period. The two conditions can be expressed as

$$\begin{cases} \delta\phi = \Delta\phi - \Delta\phi_p = 0, \\ L_1 + L_2 = \nu Z_B, \ (\nu = 1, 2, \cdots), \end{cases} \quad (7)$$

where $\Delta\phi = \phi_2 - \phi_1$, $\Delta\phi_p = \Delta k L_p$. In Eq. (7), Condition 1 indicates that the modulation phase difference in the two modulation waveguides should be compensated by the propagation phase difference of adjacent ordered modes in the unmodulated channel. Condition 2 guarantees the recovery of the Bloch momentum after two FBO processes. Only when the two conditions are simultaneously satisfied, the output spectrum will perfectly restore to its input state. For a general case, the band structures and group velocities are $k_{z,m}(k_\omega(z)) = -2C\cos[\phi_\omega(z) - \phi_m]$ and $v_{g,m}(z) = -2C\Omega\sin[\phi_\omega(z) - \phi_m]$ with $m = 1, 2$, leading to the frequency comb evolution

$$\omega(z) = \begin{cases} \omega(0) + \dfrac{2C\Omega}{\Delta k}\left[\cos(\phi_1 - \phi_0) - \cos(\Delta k z + \phi_1 - \phi_0)\right], & (0 \leq z \leq L_1) \\ \omega(L_1), & (L_1 \leq z \leq L_1 + L_p) \\ \omega(L_1) + \dfrac{2C\Omega}{\Delta k}\left[\cos(\Delta k L_1 - \Delta k L_p + \phi_2 - \phi_0) - \cos(\Delta k z - 2\Delta k L_p + \phi_2 - \phi_0)\right]. \\ & (L_1 + L_p \leq z \leq L_1 + L_p + L_2) \end{cases} \quad (8)$$

When Condition 2 is satisfied, namely, $\Delta k(L_1+L_2)=2\pi$, the output frequency center is

$$\omega_{out} = \omega(0) + \frac{8C\Omega}{\Delta k}\sin(\frac{\Delta k L_1}{2})\sin(\frac{\delta\phi}{2})\cos(\phi_1 - \phi_0 + \frac{\Delta k L_1}{2} + \frac{\delta\phi}{2}). \qquad (9)$$

If the Condition 1 is also satisfied with $\delta\phi=0$, we have $\omega_{out}=\omega(0)$. The frequency center comes back to its initial position, verifying the two conditions in Eq. (7) for spectrum reconstruction.

The above theoretical analysis can be verified by solving Eq. (3) under different input conditions. We choose $\rho_0=990$kg·m$^{-3}$ and $\beta_0=9.824\times10^{-10}$Pa$^{-1}$ for silicone rubber, such that $c_0=1014$m·s$^{-1}$ [35-37]. The modulation depth is $\Delta\beta/\beta_0=0.05<<1$, and other parameters are $\omega_0/2\pi=\omega(0)/2\pi=50$kHz, $\Omega/2\pi=200$Hz, $L_1=Z_B/3$, $L_2=2Z_B/3$ ($L_1+L_2=Z_B$) and $L_p=Z_B/3$ ($\Delta\phi_p=2\pi/3$). Figure 3(a) shows the output spectrum evolution versus the phase difference $\delta\phi=\Delta\phi-\Delta\phi_p$ under a frequency comb input $p_n(0)=\exp[-(n\Omega/W)^2]\exp(in\phi_0)$ with $W=5\Omega$ and $\phi_0=0$. The envelope manifests sinusoidal variation with respect to $\delta\phi$, which is in consistent with the theoretical prediction in Eq. (9), as also denoted by the blue solid curve in Fig. 3(a). Then we input a single frequency instead of a frequency comb into the system. As shown in Fig. 3(b), the output spectrum has a breathing pattern as $\delta\phi$ varies, which restores to a single frequency as $\delta\phi = 0$, verifying the condition in Eq. (7). Figures 3(c) illustrates the spectrum evolution for an input frequency comb under the condition of $\delta\phi = 0$. The frequency comb manifests oscillation in the encoding waveguide and keeps unchanged in the transmisission channel, and then exhibits oscillation again in the decoding waveguide, ultimately restoring to the initial input state. The envelope evolution trajectory also agrees well with theoretical prediction in Eq. (8). In Fig. 3(d), a single frequency instead of a frequency comb is incident into the system, which also restores to a single frequency at the output port though the large spectrum distortion during propagation.

More generally, since the spectrum self-imaging relies only on the conditions in Eq. (7), which is independent of the spectral functions, it should apply to an arbitrary input spectrum, either discrete or continuous ones. To verify this, we input a periodic spectrum with the frequency interval $10\Omega$ in Fig. 3(e), which can be regarded as a prototype of frequency-multiplexing communication comprising equally distributed frequency channels. The spectrum manifests channel crosstalk in the encoding waveguide and keeps unchanged in the center transmission waveguide. In the decoding waveguide, the distorted spectrum restores to its initial state through the time-reversal crosstalk. In Fig. 3(f), a continuous Gaussian spectral packet is input into the system with the width of $W=2\Omega$, which can be perfectly reconstructed at the output port. The results for other input conditions are appended in Supplementary Figs. S1 and S2 [38].

Due to the periodically oscillatory nature for FBOs [41, 42], the frequency shift is limited within the range of $|\Delta\omega|_{max}=4C\Omega/|\Delta k|$. To break this intrinsic localized feature of FBOs and realize spectrum directional transduction, we can cascade multiple FBOs with judiciously designed oscillation phases. For simplicity, we consider $M$ modulation waveguides are directly cascaded without unmodulated channels in between. The length and modulation phase of the $m$th waveguide are $L_m$ and $\phi_m$, for which the band structure and group velocity are respectively $k_{z,m}(\phi_\omega(z))=-2C\cos[\phi_\omega(z)-\phi_m]$ and $v_{g,m}(z)=-2C\Omega\sin[\phi_0-\Delta kz-\phi_m]$, ($m=1, 2, \cdots, M$). The trajectory of frequency comb evolution is

$$\omega_m(z) = \begin{cases} \omega(0)+\dfrac{2C\Omega}{\Delta k}\left[\cos(\phi_1-\phi_0)-\cos(\Delta kz+\phi_1-\phi_0)\right], & (m=1) \\ \omega(\sum_{i=1}^{m-1}L_i)+\dfrac{2C\Omega}{\Delta k}\left[\cos\left(\sum_{i=1}^{m-1}\Delta kL_i+\phi_m-\phi_0\right)-\cos(\Delta kz+\phi_m-\phi_0)\right]. & (m\geq 2) \end{cases} \quad (10)$$

To realize unidirectional spectrum transduction, we can fix each waveguide length as $L_m=Z_B/2$ and choose out-of-phase modulations $\Delta\phi=\phi_{m+1}-\phi_m=\pi$ in adjacent waveguides. For an initial phase difference $\phi_1-\phi_0=0$ or $\pi$, the frequency comb will exhibit unidirectional blue or red shift

$$\omega_m(z)=\omega(0)\pm\frac{4C\Omega(m-1)}{\Delta k}\pm\frac{2C\Omega}{\Delta k}\begin{cases}1-\cos(\Delta kz) & (m \text{ is odd}) \\ 1+\cos(\Delta kz) & (m \text{ is even})\end{cases}, \quad (11)$$

where we choose +(−) for $\phi_1-\phi_0=0(\pi)$. For both cases, the total accumulated frequency shift is $|\Delta\omega|_{max}=\omega_m(mZ_B/2)-\omega(0)=4MC\Omega/\Delta k$, which is proportional to the number of modulation waveguides. On the contrary, for $\phi_1-\phi_0=\pm\pi/2$, the frequency shift will vanish with $|\Delta\omega|=0$. For other choice of phase difference $\phi_1-\phi_0\neq 0$, $\pi$ or $\pm\pi/2$, the frequency shift satisfies $0<|\Delta\omega|<|\Delta\omega|_{max}$, which can be precisely tuned by varying the modulation phases. The theoretical analysis has also been verified by simulations in Fig. 4, where we fix $M=4$ and $L_m=Z_B/2$. Figure 4(a) shows the spectrum evolution for a frequency comb input as $\phi_1-\phi_0=0$. The spectrum experiences directional blue shift, with vanished bandwidth expansion, consistent with the theoretical trajectory denoted by the blue curve predicted by Eq. (11). While for $\phi_1-\phi_0=\pi/2$ shown in Fig. 4(b), the frequency shift vanishes and the bandwidth expansion reaches the maximum. By choosing $\phi_1-\phi_0=\pi/3$ in Fig. 4(c), both frequency shift and bandwidth expansion exist, where the amounts are less than the cases of $\phi_1-\phi_0=0$ or $\pi/2$. Finally, we input a single frequency into the system, the result shows discrete diffraction during propagation, with the bandwidth reaching the maximum of $2|\Delta\omega|_{max}=8MC\Omega/|\Delta k|$ at the output port. The effect of

directional frequency transduction also applies to the continuous spectrum, which is discussed in the Supplementary Fig. S3 [38].

In summary, we establish a framework to control sound spectrum through cascade FBOs in time-modulation waveguides systems. The wave vector mismatch in frequency transition acts as an effective constant force that drives FBOs, and the modulation phase plays the role of effective gauge potential which determines the initial oscillation phase of FBOs. Utilizing a pair of time-modulation waveguides with judiciously designed modulation phases and lengths, we demonstrate acoustic secure communication via spectrum self-imaging. By cascading multiple waveguides under out-of-phase time modulations, we realize unidirectional spectrum shift and bandwidth expansion. Finally, since all the effects do not rely on the explicit choice of incident spectrum, our work may find important applications in spectrum engineering with the operation bandwidth ranging from audible up to ultrasonic regimes.

The work is supported by the 973 Program (No. 2014CB921301), the National Natural Science Foundation of China (Nos. 11674117, 11674119, 11690030, 11690032). Y. P. acknowledges the financial support from the China Scholarship Council (CSC).

C. Q. and Y. P. contributed equally to this work.

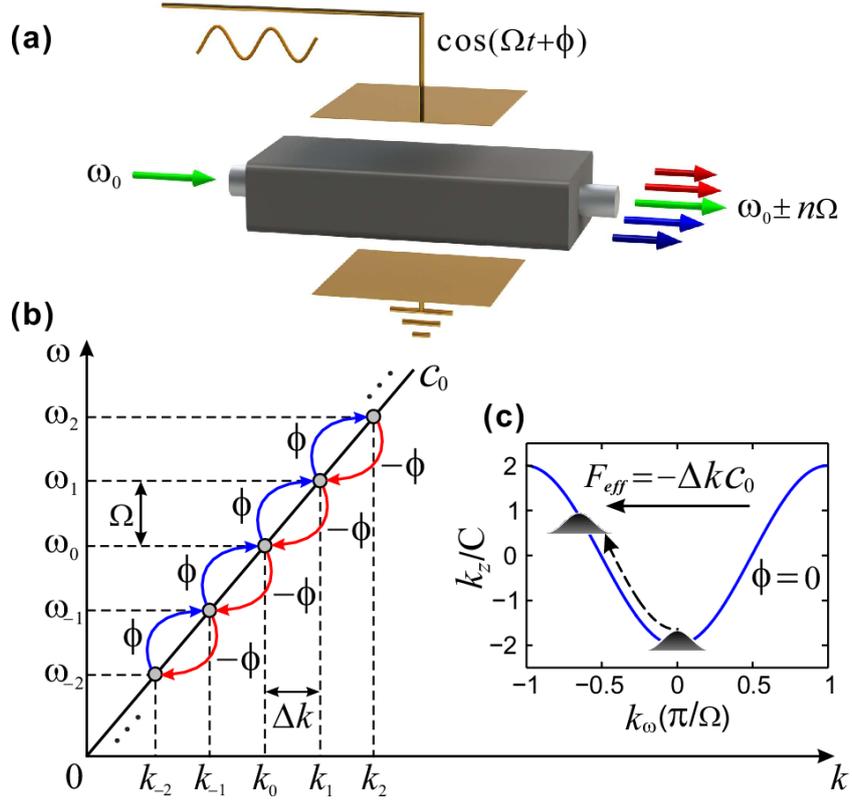

FIG. 1. (Color online) Schematic of frequency conversions in a time-modulation acoustic waveguide filled with silicone rubber. The compressibility of the filling material is subject to a sinusoidal modulation with modulation frequency $\Omega$ and initial phase $\phi$. The input and output frequencies are $\omega_0$ and $\omega_0\pm n\Omega$ ($n$=0, 1, 2, ⋯). (b) Acoustic intraband transitions in the linear waveguide band. $\Omega$ and $\Delta k$ represent the frequency and propagation constant intervals between adjacent ordered modes, with $\pm\phi$ being the phase shift in the upward and downward transitions. $c_0$ is the sound speed in the waveguide. (c) Band structure for the frequency lattice with modulation phase $\phi$=0, in which FBOs occur under the drive of an effective force of $F_{eff}=-\Delta k c_0$.

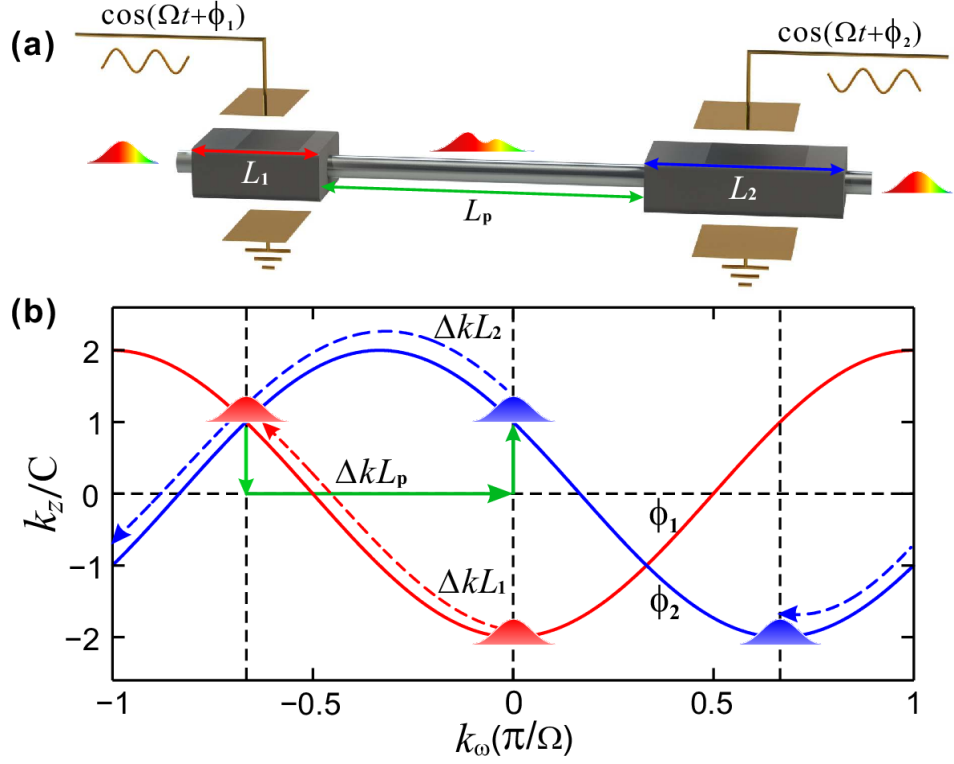

FIG. 2. (Color online) (a) A schematic diagram of two time-modulation waveguides separated by an unmodulated waveguide channel. The two waveguides with lengths of $L_1=Z_B/3$ and $L_2=2Z_B/3$ are modulated with the same frequency $\Omega$ but different phases $\phi_1=0$ and $\phi_2=2\pi/3$. The length of the unmodulated channel is $L_p = Z_B/3$, where $Z_B=2\pi/|\Delta k|$ is the Bloch oscillation period. (b) The Bloch momentum variation for an input Bloch-mode wave packet during FBOs in the two time-modulation waveguides.

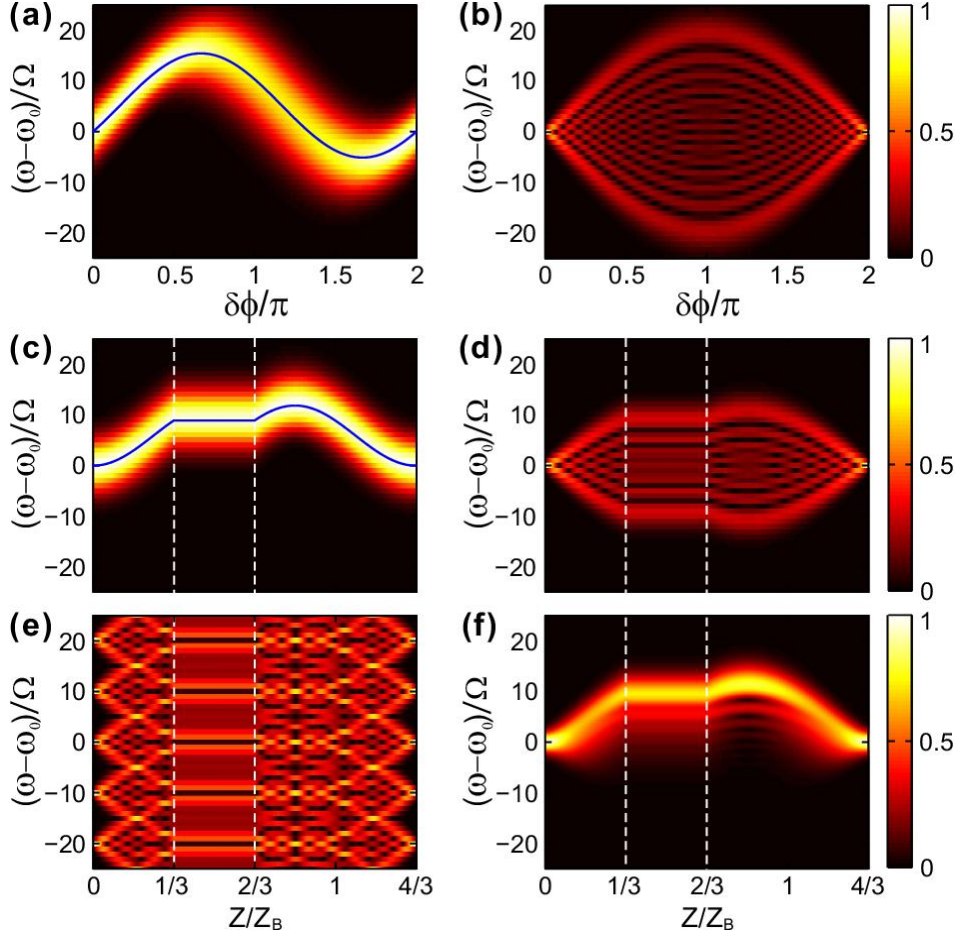

FIG. 3. (Color online) (a) Acoustic spectrum evolution versus the phase difference $\delta\phi=\Delta\phi-\Delta\phi_p$ for a frequency comb input with an initial Bloch momentum $\phi_0=0$ and comb width $W=5\Omega$. The blue curve represents the theoretical result. (b) Acoustic spectrum evolution versus $\delta\phi$ for a single frequency input. Spectrum evolutions in the proposed acoustic communication system with the input being (c) a frequency comb, (d) a single frequency, (e) a periodic discrete spectrum and (f) a continuous spectrum under the condition of $\delta\phi=0$, respectively. The blue curve in (c) denotes the theoretical trajectory of the comb envelope evolution. In (f), the width of continuous Gaussian spectrum packet is $W=2\Omega$. The white dashed lines denote the boundaries among the encoding, unmodulated, decoding waveguides.

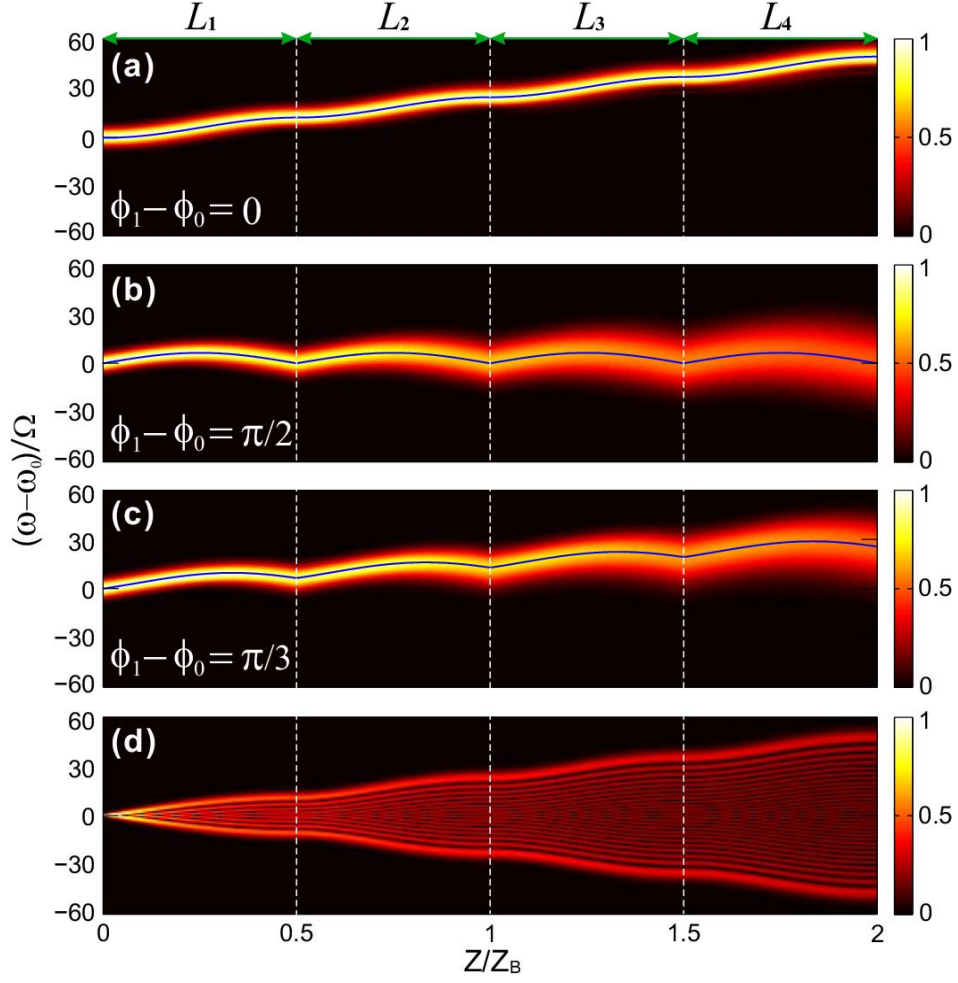

FIG. 4. (Color online) Acoustic spectrum evolutions of a frequency comb input with $\phi_0=0$ and $W=5\Omega$ in four cascading waveguides under out-of-phase time modulations. The initial phase differences are (a) $\phi_1-\phi_0 = 0$, (b) $\phi_1-\phi_0 = \pi/2$ and (c) $\phi_1-\phi_0 = \pi/3$, respectively. The length of each waveguide is fixed at $L_m=Z_B/2$ ($m=1, 2, 3, 4$). (d) Acoustic spectrum evolution for a single frequency input in the four cascading waveguides under the out-of-phase modulations. The blue curves in (a), (b) and (c) denote the trajectory of the acoustic frequency comb. The white dashed lines denote the boundaries of the four time-modulation waveguides.